\documentclass[aps, prl, twocolumn, amsmath, graphicx, latexsym, asmsymb, amsfonts]{revtex4}

\usepackage{epsfig}

\DeclareMathOperator{\tr}{tr}

\begin{document}


\title{Distillation Protocols that Involve Local Distinguishing: 
Composing Upper and Lower Bounds on Locally Accessible Information}

\author{Aditi Sen(De)$^{1,2}$,  
Ujjwal Sen$^{1,2}$, and Maciej Lewenstein\(^{2,3}\)}

\affiliation{\(^1\)ICFO-Institut de Ci\`encies Fot\`oniques, E-08860  Castelldefels (Barcelona), Spain\\
\(^2\)Institut f\"ur Theoretische Physik, Universit\"at Hannover, D-30167 Hannover, Germany\\
\(^3\)ICREA and ICFO-Institut de Ci\`encies Fot\`oniques, E-08860  Castelldefels (Barcelona), Spain
}

\begin{abstract}

We find a universal lower bound on locally accessible information for arbitrary bipartite quantum ensembles, when 
one of the parties is two-dimensional.  
In higher dimensions and in higher number of parties, the lower bound  
is on accessible information by separable operations.  
We show that for any given density matrix (of arbitrary number of parties and dimensions), there exists an ensemble, 
the 
``Scrooge ensemble'', which averages to the given density matrix and whose locally accessible information saturates the
lower bound. Moreover, we use this lower bound along with a previously obtained upper bound to obtain bounds on the yield of 
singlets in distillation protocols that involve local distinguishing.

\end{abstract}
\maketitle

Distillation of singlets from mixed states is one of the most important protocols in
viewing entanglement as a resource in quantum information \cite{NC}. It is useful in the key quantum communication tasks
 like quantum 
teleportation and quantum dense coding \cite{teleportation}. The discovery of teleportation and dense coding 
have also initiated the study of quantum channels. 
%
%
In this respect, one of the crucial questions is how much classical information 
can be encoded in an ensemble of quantum states. 
In most cases, it is quite hard to find the maximal capacity exactly, and hence the importance 
of bounds. 

There are two main purposes of this paper. 
\emph{The first purpose} 
is to 
find a universal lower bound on the maximal classical information that can be sent to two receivers, when the latter use only
local operations and classical communication (LOCC).
In a communication protocol in which a source wants to send classical information 
to a \emph{single} receiver, say Alice, by encoding it in an ensemble of quantum states, 
 important upper and lower bounds have been found on 
the maximal amount of information that  can be extracted by the receiver \cite{Holevo, JRW} 
(see also \cite{Wootters-Hulbert}).
In the case when the ensemble is sent to two separated receivers, say Alice and Bob, and whose 
task is again  to gather the maximal information, but now by using only LOCC,
an upper bound has also been given \cite{iacc-ek,iacc-dui} (see also 
\cite{Ganga-kinarey}). 
 We obtain a universal \emph{lower} bound in the case when at least one of the receivers 
is constrained to a two-dimensional 
quantum system. 
In the case when both parties are with 
higher dimensional systems, and for the case of higher number of parties,
we find the lower bound  on the maximal information attainable by separable operations between 
the receivers. 
\emph{The second purpose} of the paper is to use the lower bound for obtaining an upper bound on the yield of singlets in any distillation 
protocol that involves local distinguishing. 
We will show that indeed such bound can be obtained by using the lower bound of the present paper, 
and the upper bound of Ref. \cite{iacc-dui}, on maximal classical information obtainable by LOCC
(cf. \cite{Bimbisar-Ashok-er-dhushar-jogotey}). 
We indicate that this bound can potentially be used to detect bound entangled states with negative partial transpose \cite{dhei-ketey-dhin-dhinta}.

We note here that a two-party system, one of which is two-dimensional is of considerable practical interest, in 
other domains than considered here. 
E.g. in the ion trap quantum computer (see e.g. \cite{Inn-nodir-kinarey}), each ion has two relevant internal 
states, and is 
coupled to phonons which are described, strictly speaking by harmonic oscillators, but for any practical 
purpose, can be described by \(N\)-dimensional systems, with large \(N\).

Just as the lower bound in the case of a single receiver \cite{JRW},
 required the introduction of the concept of ``subentropy'', we will have to introduce 
the concept of ``local subentropy''.  
We also show that for any state \(\varrho\) (of any dimensions, and any number of parties), 
the so-called ``Scrooge ensemble'' \cite{JRW} corresponding to \(\varrho\), has \(\varrho\) as its 
average ensemble state, and for which the maximal information extractable by LOCC is exactly equal to 
our lower bound. We also evaluate the bound for some bipartite ensembles, for which the exact value of locally accessible information
is not known. It may be worthwhile to note here that upper bounds for the case of a single receiver \cite{Holevo}, automatically give
upper bounds for the case of multiple receivers. This however is \emph{not} true for the case of a lower bound. In particular, the 
important lower bound of Ref. \cite{JRW} for the case of a single receiver, does not give a lower bound for multiple receivers.



\emph{Accessible information for a single receiver.}
Suppose that a source encodes the information about a classical variable \(x\), that occurs with probability 
\(p_x\), in a quantum
state \(\rho_x\), and sends it to a single receiver Alice. Alice therefore receives the
 ensemble \({\cal E}= \{p_x, \rho_x\}\) from 
the source. Her task now is to gather as much information
as possible about \(x\) from the ensemble. 
Let the post-measurement ensemble, after a measurement \(M\) that gives outcome \(y\) with probability \(q_y\),
 be \({\cal E}_{out}^y = \{p_{x|y}, \rho_{x|y}\}\). The information gathered 
 by Alice from measurement \(M\) can be quantified by 
 the mutual information 
\(I_M({\cal E}:M) = H(\{p_x\}) - \sum_y q_y H(\{p_{x|y}\})\),
where \(H(\{r_i\}) = -\sum_i r_i \log_2 r_i \) is the Shannon entropy of the probability
distribution \(\{r_i\}\). Thus the  mutual
information is defined by the difference between the initial disorder and the 
disorder remaining in
the final ensemble after performing the measurement, where disorder is quantified by the Shannon
entropy. The maximal information, called the accessible information (\(I_{acc}\)), that can be gained by Alice is  
obtained by performing  maximization over all measurements: 
\(
I_{acc} = \max\limits_M I_M(x:M)
\).

\emph{Upper and lower bounds for a single receiver.}
The maximization in accessible information 
is usually hard 
to perform, 
so that it is
useful to obtain  bounds to estimate it. 
An upper bound, known as the ``Holevo bound'' \cite{Holevo}, states that
\( I_{acc} \leq \chi({\cal E}) \equiv S(\rho) - \sum_x p_x S(\rho_x)\),
where \(\rho = \sum_x p_x \rho_x\) is the average state of the ensemble \({\cal E}\), 
and \(S(\sigma) = - \tr \sigma \log_2
\sigma\) is the von Neumann entropy of \(\sigma\). 

Josza, Robb, and Wootters \cite{JRW} used the notion of subentropy, defined, for a state \(\rho\), as
\(Q(\rho) = - \sum_{k} \prod_{l\neq k} \frac{\lambda_k}{\lambda_k - \lambda_l} \lambda_k
\log_{2} \lambda_k\),
with \(\lambda_k\)s being the eigenvalues of the state \(\rho\), to obtain a lower bound:
\(I_{acc} \geq \Lambda\), where \(\Lambda({\cal E}) = Q(\rho) - \sum_i p_i Q(\rho_i)\).



\emph{Multiple receivers.}
Apart from the case of a single receiver, there are other 
important communication networks. Let us now consider a ``\(1 \rightarrow 2\) quantum network'',
where a source wants to communicate classical information to two receivers, Alice and Bob.
Suppose therefore that the source encodes the classical information \(x\) (which occurs with probability \(p_x\)
in a quantum state \(\rho_x^{AB}\) of two particles, e.g. of two photons, 
and sends the state to Alice and Bob. Alice and Bob, who are at distant locations,
obtain the ensemble \({\cal E}^{AB}= \{p_x, \rho_x^{AB}\}\).   As for the case of a single receiver, 
their aim  is  to  gather  maximal  information  about  \(x\)  by  using local quantum operations  and
classical  communication; in a similar  way, 
one  defines  the  ``locally accessible  information'' (\(I_{acc}^{LOCC}\))  
by maximizing the mutual information over 
measurements from this  restricted  class  of  operations  (LOCC):
\(I_{acc}^{LOCC} = \max I_M({\cal E}: M)\), 
where the maximization is performed over all LOCC-based measurement protocols.

 \emph{Universal upper bound on locally accessible information.} 
 Recently, 
we have shown \cite{iacc-dui}  
that for an arbitrary  given bipartite ensemble \({\cal E}^{AB}\),
that produces, after a measurement \(M\), an output (post-measurement) ensemble \({\cal E}_{out}\),
 \(I_{acc}^{LOCC} \leq  \chi_L({\cal E}^{AB})\), where
\(\chi_L({\cal E}^{AB}) \equiv S(\rho^A) + S(\rho^B) - \max\limits_{Z= A,B} \sum_x p_x 
 S(\rho_x^Z) - \overline{E}_{out}\),
 with \(\rho^{A (B)}_x = \tr_{B (A)}  \rho^{AB}_x\), 
 \(\rho^{A (B)} = \sum_x p_x \rho^{A (B)}_x\), 
and \(\overline{E}_{out}\) is an average of an arbitrary asymptotically consistent 
entanglement measure \(E\) for the output states. 
As discussed before, the Holevo bound also provides an upper bound for \(I_{acc}^{LOCC}\), because
\(I_{acc}^{LOCC} ({\cal E}^{AB}) \leq I_{acc} ({\cal E}^{AB}) \leq \chi({\cal E}^{AB})\).

\emph{Universal lower bound on locally accessible information.}
Locally accessible information  is defined as a maximization over LOCC-based measurement 
protocols, and the latter does not have a compact mathematical form. Indeed, the exact value of 
locally accessible information is known only for a very few ensembles 
(see e.g. \cite{Walgate, iacc-ek}, and references therein). 
Moreover, and in contrast to the case of the upper bound, \(\Lambda ({\cal E}^{AB})\) is not, 
in general, a lower bound for \(I_{acc}^{LOCC}\).
It is therefore extremely useful to obtain a universal lower bound, to complement the upper bound discussed above.
%
%
%

We obtain the lower bound on locally accessible information by averaging 
over all measurements on \emph{orthogonal complete pure product bases}. The main obstacle in such 
an enterprise is that the family of such bases is not well characterised at this moment. 
On the contrary, it is known that such bases can have quite nonintuitive properties. For 
example, there exists a complete orthogonal basis of pure product states, which is 
not distinguishable under LOCC
\cite{nlwe}. This 
will lead to some problems in obtaining the lower bound. 
However, at least in lower 
dimensions (precisely in \(2 \otimes n\) systems), such problems can be overcome.  

Consider therefore a measurement in the complete orthogonal pure product basis 
\(P = \{ |\alpha_j\rangle^A \otimes |\beta_k\rangle^B \} \) 
for a given ensemble \({\cal E}^{AB} = \{p_x, \rho_x^{AB}\}\).
(We will later on consider the question, whether such a measurement can actually be implemented locally).
The mutual information that is gathered  in this measurement is given by 
\(I_M({\cal E}^{AB} : P) = H(P) - H(P|{\cal E}^{AB})\),
where \(H(P)\) is the Shannon entropy of the outcome of the measurement in the basis \(P\) without 
a knowledge of the individual states of the ensemble, 
and \(H(P|{\cal E}^{AB})\) is the Shannon entropy of the outcome \emph{with} a knowledge of the same. 
%
%
Written out explicitly, 
 \(I_M({\cal E}^{AB} : P)  =  
            -   \sum_{j, k} \langle \alpha_j| \langle\beta_k| \rho^{AB} 
|\alpha_j \rangle  |\beta_k \rangle 
 \log_2   \langle \alpha_j|  \langle \beta_k| \rho^{AB} 
|\alpha_j \rangle  |\beta_k \rangle 
 +   \sum_x p_x \sum_{j, k}  \langle \alpha_j|  \langle \beta_k| \rho_{x}^{AB} 
 |\alpha_j\rangle  |\beta_k \rangle \log_2 \langle \alpha_j|  \langle \beta_k| \rho_{x}^{AB} 
 |\alpha_j\rangle  |\beta_k \rangle\), 
where \(\rho^{AB} = \sum_x p_x \rho^{AB}_x\) is the average ensemble state.

We now perform the average of \(I_M({\cal E}^{AB} : P) \)
over all 
complete orthogonal product measurements. 
After some simplification, one obtains  
\begin{equation}
\begin{array}{lcl}
 \langle I_M({\cal E}^{AB} : P) \rangle    =  
\nonumber \\
            -   d_A d_B \int d\alpha d \beta \langle \alpha| \langle\beta| \rho^{AB} 
|\alpha \rangle  |\beta \rangle 
 \log_2   \langle \alpha|  \langle \beta| \rho^{AB} 
|\alpha \rangle  |\beta \rangle  
\nonumber \\
 +  d_A d_B \sum_x p_x \int d\alpha d \beta \langle \alpha|  \langle \beta| \rho_{x}^{AB} 
 |\alpha\rangle  |\beta \rangle \log_2 \langle \alpha|  \langle \beta| \rho_{x}^{AB} 
 |\alpha\rangle  |\beta \rangle 
\nonumber \\
\equiv Q_L (\rho^{AB}) - \sum_x p_x Q_L(\rho_x^{AB}) 
\equiv \Lambda_L({\cal E}^{AB}). \nonumber 
\end{array}
\end{equation}
Here, the ensemble \({\cal E}^{AB}\)
is
from a system of dimensions \(d_A \otimes d_B\), 
and  the integrations are over all product states \(|\alpha\rangle  |\beta \rangle\). Also, 
\(Q_L(\sigma) = -   d_A d_B \int d\alpha d \beta \langle \alpha| \langle\beta| \sigma 
|\alpha \rangle  |\beta \rangle 
 \log_2   \langle \alpha|  \langle \beta| \sigma
|\alpha \rangle  |\beta \rangle \),
for a bipartite state \(\sigma\) of dimensions \(d_A \otimes d_B\).
We call \(Q_L(\sigma)\)  the ``local subentropy'' of the bipartite state \(\sigma\). Note that the 
usual subentropy of Ref. \cite{JRW} involves an integration over all orthogonal complete measurements, instead 
of the orthogonal complete \emph{product} measurements in our case. The expression for \(Q_L(\sigma)\) can be simplified 
further to give \cite{JRW, tyanara-sob-raater-dikey-asen}
\(
Q_L(\sigma) = d_A \int d\alpha Q(\langle \alpha | \sigma | \alpha \rangle) + d_A (\log_2\mbox{e})[\frac{1}{2} + \frac{1}{3} + \ldots + 
\frac{1}{d_B}]
\),
where the argument of \(Q\) is supposed to have been normalized to unit trace. 
Note that in \(2 \otimes n\) systems, the remaining integration is just over a single variable. Using this expression, 
\(\Lambda_L({\cal E}^{AB})\) takes the simple form
\begin{equation}
\label{keramoti}
\Lambda_L
= d_A  \int d\alpha \left[Q(\langle \alpha | \rho | \alpha \rangle) 
-   \sum_x p_x 
Q(\langle \alpha | \rho_x | \alpha \rangle)
 \right].
\end{equation}

Since there exists orthogonal complete product bases which cannot be exactly distinguished by LOCC,
the corresponding measurements cannot, in general, be 
implemented by LOCC \cite{nlwe}.
%
%
However, in dimensions \(2 \otimes n\) (for arbitrary \(n\)), 
there exists a simple protocol by which one can implement 
the measurement onto any orthogonal  complete product basis by LOCC \cite{UPBs} (see also \cite{Virmani-Plenio}). 
Consequently in \(2 \otimes n\), for any ensemble \({\cal E}^{AB}\), there exists at least 
one LOCC-based measurement protocol for which 
\(I_M({\cal E}^{AB} : P) = \Lambda_L({\cal E}^{AB})\), so that in general,
\begin{equation}
I_{acc}^{LOCC} \geq \Lambda_L({\cal E}^{AB}). 
\end{equation}
For the three Bell states \((|00\rangle \pm |11\rangle)/\sqrt{2}\), \((|01\rangle + |10\rangle)/\sqrt{2}\)
(with equal probabilities), 
the upper bound in Ref. \cite{iacc-ek}, and the lower bound here, give
\(0.2515 \leq I_{acc}^{LOCC} \leq 1\).

In higher dimensions (e.g. in \(3 \otimes 3\)), \(\Lambda_L({\cal E}^{AB})\) is a lower bound 
of accessible information under a larger family of operations, called the family of ``separable
superoperators''.  A separable superoperator is one which transforms bipartite states \(\sigma^{AB}\), 
defined on the Hilbert space \({\cal H}_A \otimes  {\cal H}_B\),
as
\(
\sigma \rightarrow \sum_i A_i \otimes B_i \sigma A_i^\dagger \otimes B_i^\dagger\), 
where \(A_i\) and \(B_i\) are operators on \({\cal H}_A\) and \( {\cal H}_B\) respectively, such that 
\(\sum_i  A_i^\dagger \otimes B_i^\dagger A_i \otimes B_i\) equals the identity operator on 
\({\cal H}_A \otimes  {\cal H}_B\). Note that implementation of the measurement onto 
any complete orthogonal product basis is 
a separable superoperator. In the case of higher number of parties, the generalization of the definition of \(Q_L\), that considers 
measurements on complete orthogonal product bases of all the parties, provides also a lower bound on accessible information with separable 
superoperators. 

\emph{Bound on entanglement distilable via protocols that involve distinguishing.}
We now consider distillation protocols that involve a (local) distinguishing process, which 
\emph{may or may not} correct all errors. Suppose therefore that Alice and Bob share \(m\) copies 
of the state \(\varrho^{AB}= \sum_i p_i (|\psi_i\rangle \langle \psi_i|)^{AB}\), 
where the \(|\psi_i\rangle\) may or may not be mutually orthogonal. 
Consider now the following distillation protocol. 
Alice and Bob 
share some string of the form 
\(|\psi_{i_1}\rangle \otimes \ldots \otimes |\psi_{i_m}\rangle\) with probability 
\(p_{i_1} \ldots p_{i_m}\); the corresponding ensemble being called \({\cal E}_{\varrho,m}\). 
They try to obtain the information on the string they share. This is for example the case in the 
hashing protocol \cite{Gangur-nodi-ta-kothhai}. For such protocols, we have
\(\Lambda_L({\cal E}_{\varrho,m}^{\to 2})\) \(\leq I_{acc}^{LOCC} ({\cal E}_{\varrho,m}^{\to 2}) 
\leq I_{acc}^{LOCC} ({\cal E}_{\varrho,m}) \leq \chi_L ({\cal E}_{\varrho,m})\),
where \({\cal E}_{\varrho,m}^{\to 2}\) is the ensemble \({\cal E}_{\varrho,m}\), projected on a suitably 
chosen \(2 \otimes n\) subspace. 
Considering now the entanglement measure \(E\) in \(\chi_L\) to be
the distillable entanglement, we obtain 
\[
D \leq S(\varrho^A) + S(\varrho^B) - \overline{S}_A - \Lambda_L({\cal E}_{\varrho,m}^{\to 2})/m,
\]
where \(D\) is the average entanglement distilled, per copy, in the protocol considered above, and 
\(\overline{S}_A = \sum_i  p_i S(\mbox{tr}_{B \mbox{ \footnotesize{or} } A} |\psi_i\rangle \langle \psi_i|)\).
Note that the bound is valid both in the asymptotic regime, as well as in the nonasymptotic one, with
the latter being more important in most practical applications.
For 
Bell-diagonal states (i.e. states \(\varrho\) that are diagonal in the canonical maximally entangled 
basis  \cite{ei-je-pothher-ei-dekha}) in \(d \otimes d\), we have 
\[
D \leq \log_2 d - \Lambda_L({\cal E}_{\varrho,m}^{\to 2})/m,
\]
and the result is compatible with the hashing yields \cite{{Gangur-nodi-ta-kothhai}}.

In Ref. \cite{iacc-dui}, 
we proposed a different bound on entanglement distillable in certain protocols; the one here, however,
has a larger range of applicability,
as 
the 
former
was only for distillation protocols that 
fully distinguish the strings, which e.g. requires mutually orthogonal \(|\psi_i\rangle\)'s. Even in the case of the 
hashing protocol, which uses orthogonal \(|\psi_i\rangle\)'s (the Bell states), the distinguishing is typically not complete
\cite{Gangur-nodi-ta-kothhai}.

Note that the method of obtaining the upper bound on \(D\), can be used for other future lower bounds on \(I_{acc}^{LOCC}\). 
In particular, a lower bound that can be of the order  \(\log_2 d\), can potentially be used to detect bound entangled states 
with negative partial transpose \cite{dhei-ketey-dhin-dhinta}.

\emph{Special cases.}
It is sometimes possible to further simplify the expression for \(\Lambda_L\) given in Eq. (\ref{keramoti}).
E.g., consider the case when 
the average ensemble state \(\rho^{AB}\) is a product state, i.e. it is of the form 
\(\rho^A \otimes \rho ^B\). This can happen for example in the cases when 
we want to evaluate the lower bound for a complete orthogonal  ensemble of (not necessarily product) states. 
We can then simplify the first term of \(\Lambda_{L}({\cal E}^{AB})\) (see Eq. (\ref{keramoti}))
 as
follows:   
\begin{eqnarray}
\label{simplify}
 && \Lambda_{L}({\cal E}^{AB}) \nonumber \\
 &=&  - d_A  d_B \{Q(\rho^A) + Q(\rho^B) 
 -  (\log_2\mbox{e}) [(\frac{1}{2} + \frac{1}{3} + \ldots +
\frac{1}{d_A}) 
\nonumber \\
& -&   (\frac{1}{2} + \frac{1}{3} + \ldots +
\frac{1}{d_B})]\}  + \mbox{2nd term}. 
\end{eqnarray}
This
simplification will help us to calculate the lower bound for any complete ensemble of orthogonal states.

\begin{figure}[tbp]
\begin{center}
\epsfig{figure= 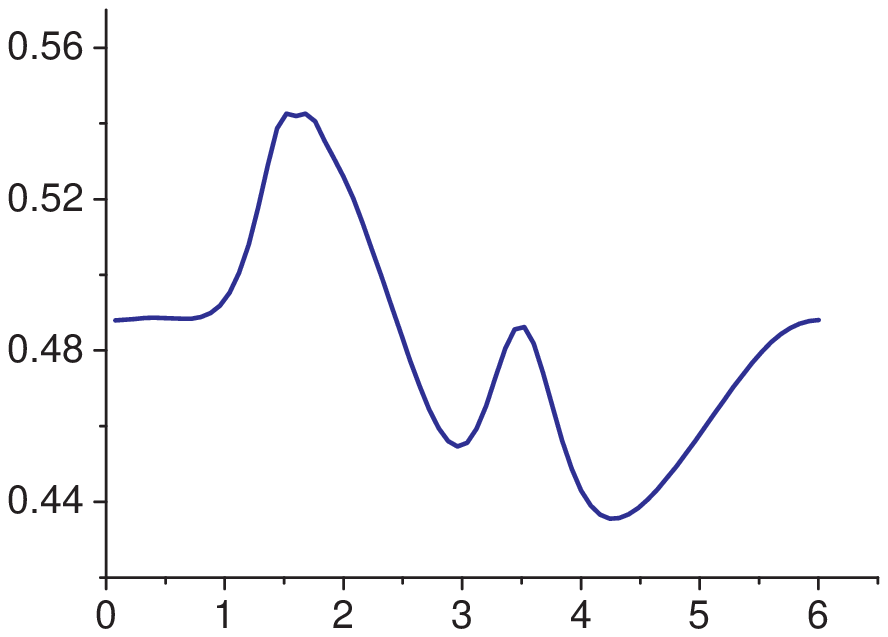,
height=.15\textheight,width=0.35\textwidth}
\put(-80,-5){\(\theta\)}
\put(-180,55){\(\Lambda_L\)}
\caption{
The lower bound for the ensemble \({\cal E}_1\). 
We set \(a = \sin \theta/2 \cos\phi/2\), \(b = \sin \theta/2 \sin \phi/2\), \(c = \cos
\theta/2\), with \(\phi = \pi/4\). 
}
\label{famousstate}
 \end{center}
\end{figure}

This happens e.g. in the case of the ensemble \({\cal E}_1\) in \(2 \otimes 2\),
 consisting of the four orthogonal states (given with equal probabilities)
\(|\psi_1\rangle  =  a |00 \rangle +  b |11 \rangle + c|10 \rangle\),
\(|\psi_2\rangle  =  k[ (b -c ) |00 \rangle +  (c -a) |11 \rangle +  (a -b)]|10 \rangle\),
\( |\psi_3\rangle  =  x |00 \rangle +  y |11 \rangle + z|10 \rangle\), 
\(|\psi_4\rangle  = |01 \rangle\), 
where \(a, b, c\) are real  numbers, 
\(k =1/ \sqrt{(b -c)^2 + (c - a)^2 + (a - b)^2}\), and \(x, y, z\) are suitable real values,
satisfying the normalization condition \(x^2 + y^2 + z^2 = 1\), and orthogonality conditions of 
\(|\psi_3\rangle \) with \(|\psi_1 \rangle\) and 
\(|\psi_2 \rangle\).
The exact locally accessible information for this ensemble is not known. 
Again, the average state for this ensemble is
the identity in the four dimensional complex Hilbert space, and hence is of the form \(\rho^A \otimes
\rho^B\). 
In Fig. \ref{famousstate}, we draw the lower
bound on locally accessible information with respect to a single parameter (using Eq. (\ref{simplify})). 


 If we have an ensemble consisting of only pure product states (not neccessarily orthogonal),
  and for which 
the average ensemble state is 
of the form \(\rho^A \otimes \rho^B\), 
 the lower bound can be further simplified
as 
\( \Lambda_L = Q(\rho^A) + Q(\rho^B)\).
An example 
of this situation is 
an 
ensemble  consisting of 
\(|00\rangle, |01\rangle, |10\rangle, |11\rangle, |++\rangle, |+-\rangle, |-+\rangle, |--\rangle\)
(with equal probabilities), where
\(|\pm\rangle = (|0\rangle \pm |1\rangle)/\sqrt{2}\).

\emph{Saturation.}
We now show that for any given multiparty density matrix \(\rho\) (of any dimensions), there exists an ensemble (called the Scrooge 
ensemble \cite{JRW}), for which the average state is \(\rho\), and 
\(I_{acc}^{LOCC} = \Lambda_L\).
Let \(\{|e_i\rangle\}_{i=1}^{N}\) (\(\{\lambda_i\}_{i=1}^{N}\)) be the eigenbasis
(eigenvalues) of \(\rho\).
The eigenbasis may 
contain 
entangled states.
Then the Scrooge ensemble \({\cal E}_{\cal S}\) is the (continuous) distribution of \(|\{x_i\}\rangle = \sum_i \sqrt{x_i} |e_i\rangle\),
distributed as \((n-1)! n dx_1 \ldots dx_{N-1}/[\lambda_1  \ldots \lambda_{N-1} (x_1/\lambda_1 + \ldots + x_N/\lambda_N)^{N+1}]\).
It was shown in Ref. \cite{JRW} that the amount of mutual information that is obtained in a complete orthogonal measurement on  
\({\cal E}_{\cal S}\)
is a constant. 
Therefore, 
the mutual information obtained by measuring onto any \emph{multi-orthogonal} complete product basis will 
be a constant 
(\(= Q_L ({\cal E}_{\cal S})\)), and such measurement can be performed locally for any dimensions and any number of parties. 
Finally, the maximal mutual information
for \emph{global} measurements is, in general, attainable on complete measurements \cite{eta-holo-Davies}, so that for \({\cal E}_{\cal S}\),
\(I_{acc}^{LOCC} = Q_L\) (because \(I_{acc}^{LOCC}\) is sandwiched between \(I_{acc}\) and \(Q_L\), in this case).
To our knowledge, this is the only known globally 
indistinguishable ensemble for which \(I_{acc} = I_{acc}^{LOCC}\).

\emph{Summary.}
We have obtained a universal lower bound on locally accessible information for arbitrary bipartite 
ensembles, in the case when one of the parties has a two-dimensional system. For higher dimensional
and multipartite systems, 
the universal bound is for separable operations. We have shown that for any given 
multiparty state \(\rho\)
there exists an ensemble
whose local accessible information  saturates our bound, and whose average state is \(\rho\).
We use this lower bound along with a previously obtained upper bound on the same quantity, to 
give an upper bound on the yield of singlets in distillation protocols that involve local distinguishing.



We acknowledge support from the 
DFG (SFB 407, SPP 1078, SPP 1116, 436POL), 
the Alexander von Humboldt Foundation, the EC Program 
QUPRODIS, the ESF Program QUDEDIS, and EU IP SCALA.


\begin{thebibliography}{10}
\bibitem{ei-je-pothher-ei-dekha} The canonical maximally entangled basis in \(d \otimes d\) are 
\(\frac{1}{\sqrt{2}}\sum_{i=0}^{d-1}e^{2\pi i j n /d} |j\rangle |(j+m)\mbox{mod }d \rangle\), (\(n,m=0, \ldots, d-1\)).
\end{thebibliography}
\end{document}